\pdfoutput=1
\documentclass[11pt]{article}

\usepackage{acl}

\usepackage{times}
\usepackage{latexsym}
\usepackage{adjustbox} 
\usepackage{times}
\usepackage{latexsym}
\usepackage{amsmath}
\usepackage{multirow}
\usepackage{graphicx}
\usepackage{booktabs}
\usepackage{amssymb}
\usepackage{mathtools}
\usepackage{cite}
\usepackage{url}

\usepackage[T1]{fontenc}
\usepackage[utf8]{inputenc}

\usepackage{microtype}

\usepackage{inconsolata}

\title{
  \includegraphics[width=2cm]{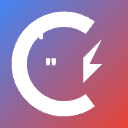} \\[1ex]
  CiteClick: A Browser Extension for Real-Time Scholar Citation Tracking
}

 \author{Nishat Raihan \\ George Mason University, USA \\
 \texttt{mraihan2@gmu.edu}
 }
 
\begin{document}
\maketitle
\begin{abstract}
This technical report presents CiteClick \footnote{\url{github.com/mraihan-gmu/CiteClick/tree/main}}, a browser extension designed to monitor and track Google Scholar citation counts for multiple researchers in real-time. We discuss the motivation behind the tool, its key features, implementation details, and potential impact on the academic community. The report covers installation procedures, usage guidelines, and customization options, concluding with a discussion on future work and potential improvements. By automating the process of citation tracking, CiteClick aims to enhance research evaluation processes and facilitate more informed decision-making in academic contexts.
\end{abstract}

\section{Introduction}
In the academic world, citation counts serve as a crucial metric for assessing the impact and reach of scholarly work \citep{hirsch2005index}. The h-index, introduced by Hirsch, has become a standard measure of a researcher's productivity and impact, relying heavily on accurate citation counts. However, manually tracking citation counts for multiple researchers can be time-consuming and inefficient, particularly given the rapid pace of modern academic publishing \citep{meho2007rise}.

To address this issue, we present CiteClick, a browser extension that automates the process of monitoring Google Scholar citation counts for multiple researchers. By leveraging the accessibility of Google Scholar and the convenience of browser extensions, CiteClick provides a user-friendly solution to the challenge of citation tracking. This tool has the potential to significantly streamline research evaluation processes and provide valuable insights into the evolving landscape of academic impact.

\section{Related Work}

The development of CiteClick builds upon a rich body of work in the fields of bibliometrics, academic impact assessment, and research productivity tools. Several existing solutions and research efforts have paved the way for tools like CiteClick:

\textbf{Citation Analysis Tools}: Platforms like Publish or Perish \citep{harzing2007publish} have long provided desktop applications for citation analysis. Unlike CiteClick, these tools often require manual data input and lack real-time updating capabilities.
    
\textbf{Academic Social Networks}: Websites such as ResearchGate and Academia.edu \citep{thelwall2014researcher} offer citation tracking as part of their broader academic networking features. However, these platforms rely on user-uploaded content and may not provide comprehensive coverage of all publications.
    
\textbf{Database-Specific Tools}: Web of Science and Scopus provide their own citation tracking tools \citep{mongeon2016journal}, but these are limited to publications within their respective databases and often require institutional subscriptions.
    
\textbf{Altmetrics}: Tools like Altmetric.com \citep{priem2010altmetrics} track alternative measures of impact, including social media mentions and policy citations. While providing valuable complementary data, these tools do not focus on traditional citation counts.
    
\textbf{Browser Extensions for Academics}: Extensions like Lazy Scholar \citep{ordunamaleadesign} enhance Google Scholar search results with additional metadata but do not provide the multi-researcher tracking capabilities of CiteClick.

CiteClick differentiates itself by combining the ease of use of a browser extension with real-time, multi-researcher citation tracking specifically from Google Scholar. This approach fills a gap in the existing landscape of citation analysis tools, offering a balance between comprehensive coverage and user-friendly, on-demand access to citation data.

\section{Features and Implementation}
CiteClick offers a comprehensive set of features designed to enhance the citation tracking experience. These features, along with their implementation details, are as follows:

\begin{itemize}
    \item \textbf{Real-time citation counts}: CiteClick fetches up-to-date data from Google Scholar, ensuring users have access to the most current information. This is crucial in the fast-paced world of academic publishing \citep{bar2008use}. The extension uses asynchronous JavaScript to send requests to Google Scholar, parse the HTML response, and extract the relevant citation data.
    
    \item \textbf{Multi-researcher tracking}: Users can monitor citation counts for multiple scholars simultaneously, facilitating comparative analysis. This feature is implemented through a configurable list of Google Scholar IDs, stored in the extension's local storage for quick access and modification.
    
    \item \textbf{Dynamic ranking system}: CiteClick automatically sorts tracked researchers based on their citation counts. This feature enables users to quickly identify high-impact researchers and track changes in relative standings over time \citep{ruane2017bibliometrics}. The ranking algorithm is implemented in JavaScript and updates in real-time as citation data is refreshed.
    
    \item \textbf{One-click updates}: Users can refresh all citation data with a single click, streamlining the update process. This is achieved through a background script that manages data fetching and storage operations \citep{flanagan2020javascript}.
    
    \item \textbf{Local storage}: The extension utilizes browser local storage for offline access and improved performance \citep{zakas2010high}. This ensures that users can access their tracked researchers and the most recently fetched citation data even without an internet connection.
\end{itemize}

From a technical perspective, CiteClick is implemented as a browser extension using modern web technologies. The core components include:

\begin{itemize}
    \item \textbf{Manifest file}: This JSON file defines the extension's properties, permissions, and structure \citep{browser2021manifest}. It specifies the extension's name, version, description, and the scripts and resources it uses.
    
    \item \textbf{Background script}: This JavaScript file runs continuously in the background, handling data fetching and storage operations. It uses the Fetch API to make requests to Google Scholar and the chrome.storage API for data persistence.
    
    \item \textbf{Popup interface}: This HTML and JavaScript combination provides a user-friendly means of viewing and updating citation data \citep{haverbeke2018eloquent}. It uses modern DOM manipulation techniques to dynamically update the UI based on the stored citation data.
    
    \item \textbf{Configuration file}: This JavaScript file allows users to customize the list of tracked researchers. It exports a configuration object that can be imported and used by other parts of the extension.
\end{itemize}

The extension's architecture ensures efficient performance and minimal impact on browser responsiveness \citep{archibald2018using}. By leveraging asynchronous operations and local storage, CiteClick can provide a smooth user experience even when dealing with large numbers of tracked researchers or slow network conditions.

\section{Installation and Usage}
The installation process for CiteClick follows standard practices for loading unpacked browser extensions \citep{google2021loading}. The steps are as follows:

\begin{enumerate}
    \item Download the latest release from the project repository.
    \item Unzip the downloaded file to a local directory.
    \item Navigate to the browser's extension management page (e.g., chrome://extensions for Chrome).
    \item Enable "Developer mode" to allow the loading of unpacked extensions.
    \item Select "Load unpacked" and choose the directory containing the unzipped CiteClick files.
\end{enumerate}

Once installed, CiteClick can be accessed via its icon in the browser toolbar. The popup interface displays a list of tracked researchers along with their current citation counts. Users can update the data by clicking the "Update Citations" button, triggering a refresh of all tracked researchers' information.

Customization is a key feature of CiteClick, achieved through modification of the `config.js` file. This file allows users to specify the Google Scholar IDs of researchers they wish to track. The process for adding a new researcher is as follows:

\begin{enumerate}
    \item Locate the researcher's Google Scholar profile.
    \item Extract the ID from the URL (e.g., 'vAx7VsoAAAAJ' from \url{https://scholar.google.com/citations?user=vAx7VsoAAAAJ}).
    \item Open the `config.js` file in a text editor.
    \item Add the new ID to the `scholarIds` array.
    \item Save the file and reload the extension in the browser.
\end{enumerate}

This flexibility enables the extension to be tailored to individual research interests or institutional needs \citep{mcfedries2010app}. Users can easily add or remove researchers from their tracking list, ensuring that the tool remains relevant and useful as research focuses evolve.

\section{Future Work and Conclusion}
While CiteClick represents a significant step forward in automating citation tracking, there are several avenues for future development and enhancement:

\begin{itemize}
    \item \textbf{Integration with other databases}: Expanding CiteClick to include data from platforms such as Scopus or Web of Science could provide a more comprehensive view of citation patterns across different platforms \citep{martin2018google}.
    
    \item \textbf{Data visualization}: Implementing tools to represent citation trends over time could offer valuable insights into the trajectory of a researcher's impact \citep{borner2010atlas}. This could include line graphs, bar charts, or even more advanced visualizations like network graphs of co-citations.
    
    \item \textbf{Alert systems}: Developing notification mechanisms to alert users of significant changes in citation counts could enhance the tool's proactive capabilities \citep{priem2010altmetrics}. This could involve push notifications, email alerts, or in-browser notifications.
    
    \item \textbf{Server-side component}: To improve scalability and enable cross-device synchronization, introducing a server-side component could be considered. This would reduce the load on individual browsers and facilitate a more seamless user experience across multiple devices \citep{taivalsaari2011web}.
    
    \item \textbf{Machine learning integration}: Incorporating machine learning algorithms could enable predictive analytics, such as forecasting future citation trends or identifying emerging influential researchers in specific fields.
\end{itemize}

In conclusion, CiteClick represents a valuable contribution to the field of academic impact assessment. By providing real-time, easily accessible citation data for multiple scholars, it has the potential to streamline research evaluation processes and facilitate more informed decision-making in academic contexts. As the tool evolves and incorporates user feedback, it is poised to become an indispensable resource for researchers, administrators, and anyone interested in tracking the impact of scholarly work.

The development of CiteClick opens up new possibilities for understanding and quantifying academic influence. Future work will focus on expanding its capabilities, improving its user interface, and integrating more advanced features to meet the evolving needs of the academic community. As bibliometrics continue to play a crucial role in research assessment, tools like CiteClick will be essential in providing accurate, up-to-date, and easily accessible citation data.

\section{Conclusion and Future Directions}

CiteClick represents a significant advancement in the realm of citation tracking tools, offering researchers and academic administrators a streamlined, real-time solution for monitoring scholarly impact. By leveraging the accessibility of Google Scholar and the convenience of browser extensions, CiteClick addresses the growing need for efficient, up-to-date citation data in the increasingly competitive academic landscape.

The key contributions of CiteClick include:

\begin{itemize}
    \item Real-time, multi-researcher citation tracking directly from Google Scholar.
    \item A user-friendly interface that simplifies the process of monitoring academic impact.
    \item Local storage capabilities that enable offline access to citation data.
    \item A flexible configuration system that allows users to customize their tracking lists.
\end{itemize}

As bibliometrics continue to play a crucial role in research assessment and career advancement, tools like CiteClick will become increasingly valuable. The ability to quickly and easily access citation data for multiple researchers can inform decision-making processes in hiring, promotion, and funding allocation, as well as help individual researchers track their own impact and identify potential collaborators.

Future work on CiteClick will focus on several key areas:

\begin{enumerate}
    \item \textbf{Enhanced Data Integration}: Expanding CiteClick to incorporate data from additional sources such as Scopus and Web of Science will provide a more comprehensive view of a researcher's impact across different platforms.
    
    \item \textbf{Advanced Visualization Tools}: Implementing interactive graphs and charts will allow users to visualize citation trends over time, offering deeper insights into the trajectory of a researcher's influence.
    
    \item \textbf{Predictive Analytics}: Incorporating machine learning algorithms could enable CiteClick to forecast future citation trends and identify emerging influential researchers in specific fields.
    
    \item \textbf{Collaboration Features}: Adding the ability to share and compare citation lists could facilitate collaboration and benchmarking among research groups or institutions.
    
    \item \textbf{Integration with Research Management Systems}: Developing APIs to allow CiteClick to integrate with institutional research management systems could streamline administrative processes and provide more comprehensive research analytics.
\end{enumerate}

As CiteClick evolves, it has the potential to become an indispensable tool in the academic ecosystem, facilitating more informed, data-driven decision-making in research evaluation and career development. By continuing to refine its features and expand its capabilities, CiteClick aims to meet the ever-changing needs of the academic community and contribute to a more transparent and efficient system of scholarly impact assessment.

\bibliography{references}

\end{document}